# Small-Angle X-ray and neutron scattering from diamond single crystals

**A A Shiryaev[1], P Boesecke[2]**

[1]Institute of Physical Chemistry and Electrochemistry RAS, Leninsky pr. 31, 1199991, Moscow, Russia

[2] ESRF, 6 rue Jules Horowitz, BP220, 38043 Grenoble Cedex, France

E-mail: shiryaev@phyche.ac.ru

**Abstract**. Results of Small-Angle Scattering study of diamonds with various types of point and extended defects and different degrees of annealing are presented. It is shown that thermal annealing and/or mechanical deformation cause formation of nanosized planar and three-dimensional defects giving rise to Small-Angle Scattering. The defects are often facetted by crystallographic planes 111, 100, 110, 311, 211 common for diamond. The scattering defects likely consist of clusters of intrinsic and impurity-related defects; boundaries of mechanical twins also contribute to the SAS signal. There is no clear correlation between concentration of nitrogen impurity and intensity of the scattering.

## 1. Introduction

Small-angle X-ray (SAXS) and neutron (SANS) scattering are well-known methods of investigation of nanosized heterogeneities in solids and is widely employed for studies of aging of carious compounds and alloys. However, systematic studies of defects in diamonds are lacking. The first significant SAS study of synthetic diamonds using SAXS is described in [1]. It was shown that some synthetic diamonds possess anisotropic scattering ascribed to nanosized metal lamellae, entrapped during crystal growth. However, the progress in diamond growth under High-Pressures – High-Temperatures (HPHT) conditions is remarkable and modern crystals are virtually free from macro- and microinclusions (e.g., [2]). In several papers [3-8] we have demonstrated that prior to thermal annealing diamonds belonging to types IIa (nitrogen-free) and Ib (with single substitutional N atoms) show negligible intensity of small-angle scattering, but annealing and/or deformation of diamonds at HPHT-conditions lead to appearance of the scattering without relationship to presence of nitrogen defects in the lattice. Intensity and symmetry of the scattering depend on conditions of the sample treatment. The scattering is caused by defects with sizes in the range from 8-9 till ~60 nm; some of the defects are oriented in diamond lattice and are thus facetted or planar. Some of these defects, likely, influence thermal conductivity and mechanical properties of diamonds [9]. There is no unique correlation between nitrogen concentration in diamond lattice and the SAS properties, but a general trend is clear: the size of the scatterers decreases and their contrast with the diamond matrix increases with advancement of the annealing. The scattering from diamonds with low or moderate degree of nitrogen aggregation is generally weak and is due to large defects with broad size distribution. The scattering from heavily annealed diamonds of IaB type is strong and is probably related to voidites-like defects.

In this paper we present results of investigation of natural and synthetic diamonds using Small-Angle X-ray Scattering. This paper extends results obtained using neutron scattering for the same set of diamonds [6].

## 2. Experimental

Natural and synthetic diamonds with various concentrations of the principal chemical impurity - nitrogen – were studied. Concentration and type of N-defects were estimated using standard approaches by employing Infra-red spectroscopy [9-11]. Some of the specimens preserved crystallographic faceting; some were represented by flat plates with (100) orientation laser cut through the crystal center. Conditions of HPHT deformation and annealing as well as information about microstructure of the samples are described in detail in [7].

Small-angle scattering was studied at beamline ID01 at ESRF, Grenoble. The sample-detector distance was 4192 cm, the scattering path was under vacuum. Raw experimental data were treated using software described in [12]; calibration of the scattering intensity was performed using a Lupolen standard. Size distribution of the scatterers was calculated using GNOM software [13].

Already in early papers on SAXS studies of irradiated diamonds [14] it was shown that in case of certain orientation of the crystal towards the incoming beam double Bragg scattering (DBS, see details in [15]) may appear. An example of the SAXS pattern with significant contribution of the double scattering is shown on figure 1. It is obvious that interpretation of such patterns is very problematic. To avoid the double scattering the measurements were performed at several different energies around 8 keV. Additional control of the potential presence of the DBS was performed by comparison of anisotropies of X-ray and neutron scattering patterns. In the SANS experiments the DBS effect was fully absent since the employed wavelength (6 Å [6]), is more than two times larger than the largest interplanar distance in diamond ($d_{111}$ = 2.06 Å).

It is important to note that qualitatively the 2D SAXS and SANS patterns are fairly similar, but the statistics of the X-ray data is much higher. In addition, in the X-ray experiments averaging of the scattering curves obtained at several energies was performed. Therefore, the main part of quantitative results was obtained from the SAXS measurements.

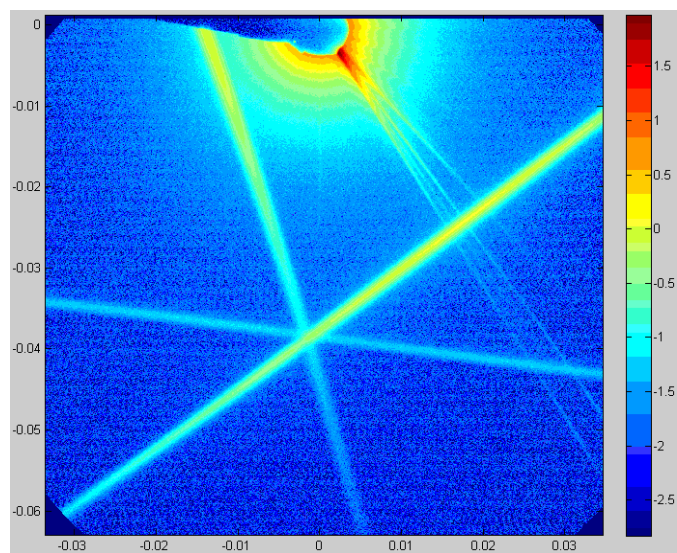

Figure 1. An example of 2D SAXS pattern with pronounced contribution of Double Bragg Scattering

## 3. Results

The small-angle scattering patterns from the majority of the studied diamonds with moderate degrees of nitrogen aggregation (i.e., with moderate post-growth annealing) are usually isotropic. This indicates that the shape of the scatterers is rather isometric. In principle, the isometric character of the scattering could result from random mutual orientation of the incident X-ray beam and anisotropic particles as it happens in solutions. However, our dedicated experiments demonstrate that in diamonds the anisotropy of the SAS patterns, if present at all, is pronounced even when the misalignment is significant.

Typical SAXS curves for moderately annealed diamonds are shown at figure 2a. The size distribution of the spherical scatterers is shown on figure 2b. In such diamonds the observed defects are heavily polydisperse with relatively abundant defects with radii between 50 and 100 nm. These results are in accordance with our previous studies of diamonds with low and moderate degrees of nitrogen aggregation [3-5, 8].

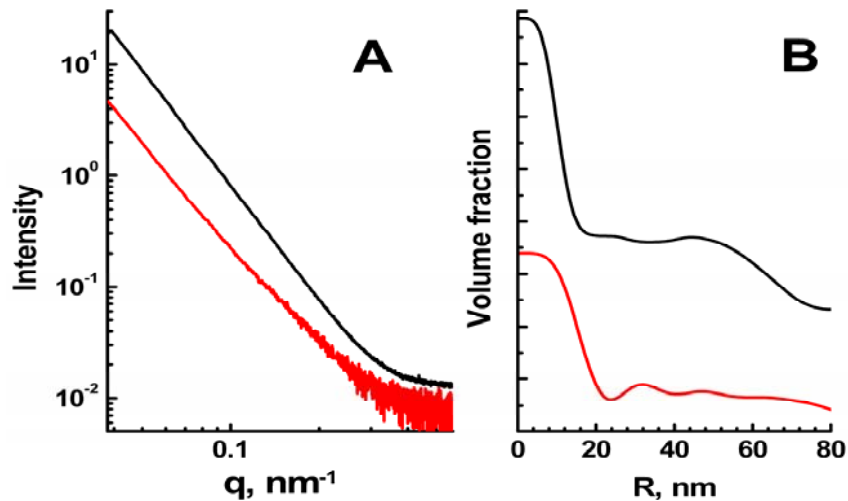

Figure 2. SAXS curves and (A) and size distribution of the scatterers
in diamonds with moderate degree of nitrogen aggregation

Small-angle scattering from some specimens is markedly anisotropic. In most cases such diamonds are either deeply annealed in nature (diamonds of pure IaB type) or heavily deformed mechanically. For the IaB diamond (580 at. ppm of nitrogen) the X-ray and neutron scattering is strong and markedly anisotropic (figure 3a). The "rays" of the scattering patterns are directed along directions [110], [111], [311]. The scattering curves along these directions are shown on figure 3b. Broad Bragg peaks along the [111] indicate ordering of the scattering defects with the period approx. 13.6 nm. Moreover, weak peaks with periodicities approx. 25 and 36 nm are also present. Taking into account their considerable width and uncertainty in determination of the peak positions, it seems likely that these peaks correspond to doubled and tripled period of 13.6 nm. The existence of the periodicity in spatial distribution of the scatterers is rather unusual. The gyration radii are equal to 36 nm in [111] and 25 nm in [110]. The slope of the scattering curve in [110] on double logarithmic plot suggests that the shape of the scatterers in this direction is intermediate between disks and cylinders, but due to obvious polydispersity of the system such interpretation should be treated with caution.

One of the specimens was mechanically deformed at HPHT conditions (6.5 GPa, 1600 °C); a mixture of SiC powder and hard silicates was employed as a pressure transmitting medium in direct contact with the diamond sample. Transmission electron microscopy study of this sample showed that the deformation has produced high dislocation density with certain degree of polygonisation as well as mechanical twins [7]. The SAS patterns of this diamond are markedly anisotropic. The gyration radii of the scatterers are similar for different directions and are between 32 and 36 nm. The most remarkable peculiarity of this sample is the existence of relatively weak Bragg peaks in several crystallographic directions (figure 4). In [111] the peak (probably an unresolved doublet) corresponds to the period 23-26 nm; possibly doubled period (46-47 nm) also exists. The curve in [110] contains numerous peaks with periodicities between 14.5 and 23 and 65-77 nm.

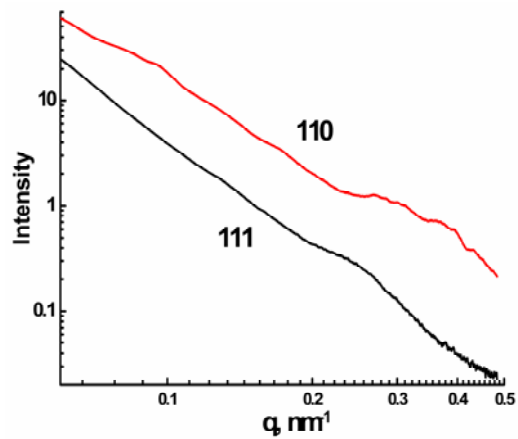

Figure 4. Deformed diamond: SAXS curves along different crystallographic directions.

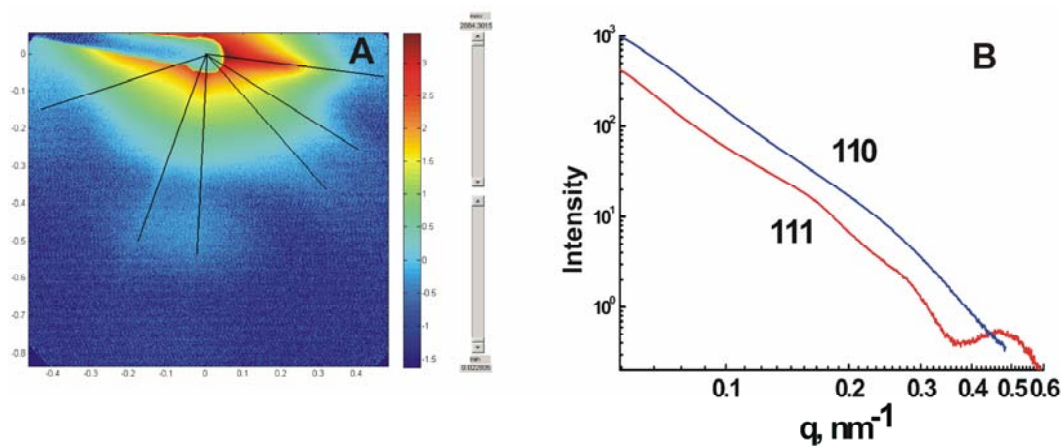

Figure 3. 2D SAXS pattern (A) the scattering curves (b) along two crystallographic directions for the type IaB diamond (the highest N aggregation state).

One of the studied samples - BR2 – belongs to so-called fibrous diamonds. Natural diamonds of this variety contain numerous submicroscopic inclusions of the growth medium and are of considerable interest for geosciences [9, 16]. The scattering pattern from this sample is obviously anisotropic and is characterised by eight-fold symmetry (figure 5). The scatterers are bound by (100) and (110) planes. Most likely, the scattering is mostly caused by facetted microinclusions. Existence of several subpopulations of the microinclusions - faceted oriented and irregular - in fibrous diamonds was discovered in TEM study [17]. These subpopulations differ in composition and origin. Large slope of the scattering curve on a log-log plot (~-5) indicates that the sizes of the scatterers exceed the resolution of the employed setup.

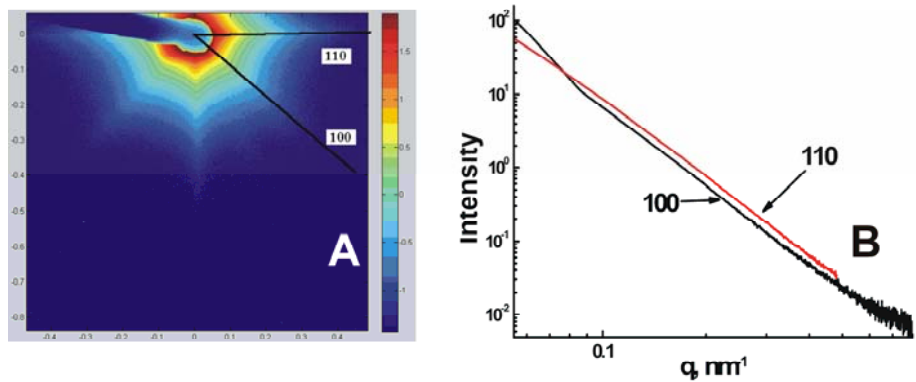

Figure 5. 2D SAXS pattern (A) and scattering curves (B) for a fibrous diamond.

## 4. Discussion

Results of the current study support conclusion of our previous works that many diamond crystals contain defects with sizes up to several tens nanometers which are formed as a result of diffusion of intrinsic and extrinsic defects during high temperature annealing. The SAXS and SANS [6] results also show that mechanical deformation also leads to formation of strongly scattering defects. Application of two-dimensional detectors reveals that some of the defects responsible for the SAS are highly anisotropic. The scattering anisotropy indicates that the scatterers produced during long-term annealing or mechanical deformation are facetted and/or highly anisotropic. The preferred directions – [100], [110], [111], and [113] – correspond to the most important crystallographic axes in diamond lattice. The faceting of macroscopic inclusions of foreign phases in diamond is a well-known phenomenon. Direct microscopic information about nanosized inclusions in diamonds is limited, but data on voidites shows that the majority of these defects are bounded by (111) planes. Most likely, this indicates achievement of an equilibrium shape. The SAS data shows that nanosized defects in diamonds could be facetted not only by common octahedral and cubic planes, but also by (110) and (113). This variability can be explained by changes in chemical composition of the trapped matter and corresponding changes in wettability of the diamond planes.

Most likely, voidites are the principal scatterers in heavily annealed diamonds. These defects are characterized by significant contrast of electron density with the diamond matrix. The SAS signal from type IaB diamonds, which usually contain voidites, is strong; therefore, SAS techniques represent a method of express selection of voidites-rich stones for subsequent detailed studies of voidites. Chemical composition of the scatterers in diamonds with moderate degrees on annealing remains poorly defined. Their sizes are considerable (tens of nanometers) and the contrast with the matrix should be small, since the SAS intensity is low-to-moderate. SANS and SAXS data [4, 6] show that even annealing of nitrogen-free diamonds leads to appearance of the scattering. Therefore, presence of chemical impurities such as nitrogen is not necessary for formation of the scatterers. Presumably, clusters of vacancies and interstitials may also give rise to the SAS signal. Existence of such clusters is indirectly supported by diffuse X-ray scattering [19], electron microscopy [20] and positron annihilation spectroscopy [21].

We also observe ordering in the spatial distribution of the extended defects. Similar ordering is often observed in aged alloys and semiconductors (e.g., [22]). However, in aged metal alloys the ordering results from high concentrations (several at.%) of chemical impurities, segregating into the new phases. Concentration of impurities in diamonds is always much smaller. It is thus not very probable that the ordered scattering defects fill the whole volume of a diamond crystal. More plausible model implies existence of localized lattice volumes with high concentration of the defects. As examples of such situation one may recall that voidites are usually concentrated around degraded platelets [18] or on dislocations [23]. Such localization can be explained by directed diffusion of intrinsic and extrinsic point defects in strain fields surrounding the extended defects and by decay of supersaturated solid solution.

What is the nature of the scatterers produced by mechanical deformation? We have earlier demonstrated that microstructure of deformed diamonds depends on the treatment conditions and the deformation may lead to mechanical twinning or to appearance of dislocations with variable degrees of polygonisation. Mechanical twinning of diamond lattice may create lamellae of other (hexagonal) diamond polytypes. The differences in electron density of the diamond polytypes is small, but certain configurations of the interpolytype boundary may give rise to strong anisotropic Small-Angle Scattering.

## 5. Acknowledgments
The work was partially supported by Alexander von Humboldt foundation (AAS). We thank Drs. V.V. Volkov and E. Stoyanov for help in the experiments and for discussions.